\documentclass[11pt,a4paper,onecolumn,notitlepage]{article}

\usepackage{CJKutf8}
\usepackage{cmap} 
\usepackage[pdfpagemode=None,
pagebackref=true,
bookmarksopen=false,
hyperindex=true,
colorlinks=true,
linkcolor=blue,
citecolor=blue,
pdftitle={Generalised Br\`{e}gman relative entropies: a brief introduction},
pdfauthor={Ryszard Pawe{\l} Kostecki}]{hyperref}

\usepackage{tabularx} 
\usepackage{amsfonts} 
\usepackage{amsmath} 
\usepackage{mathrsfs} 
\usepackage{amssymb} 
\usepackage[OT2,T1]{fontenc} 
\usepackage[2cell,all]{xy} 
\usepackage{graphicx} 
\usepackage{bm}

\DeclareMathAlphabet{\mathpzc}{OT1}{pzc}{m}{it}

\usepackage{geometry}
\usepackage[toc]{multitoc}
\usepackage{bibspacing}
\usepackage{cyrillic}
\usepackage{rpk}

\usepackage{empheq}
\usepackage{setspace} 
\newcommand{\cyrrm}[1]{\mbox{\fontencoding{OT2}\fontfamily{wncyr}\selectfont#1}} 

\renewcommand*{\backref}[1]{}
\renewcommand*{\backrefalt}[4]{%
  \ifcase #1 %
    \relax
  \or
    $\uparrow$~#2.
  \else
    $\uparrow$~#2.
  \fi%
}

\begin{document}
\title{{\vskip -1.5cm}{\LARGE Generalised Br\`{e}gman relative entropies: a brief introduction}}
%
%
\author{Ryszard Pawe{\l} Kostecki\\
{\small\textit{International Center for Theory of Quantum Technologies, University of Gda\'{n}sk}}\\
{\small\textit{ul. Jana Ba\.{z}y\'{n}skiego 1a, 80--309 Gda\'{n}sk, Poland}}\\
{\small\texttt{kostecki@fuw.edu.pl}}\\
{\small\href{https://www.fuw.edu.pl/~kostecki}{https://www.fuw.edu.pl/$\sim$kostecki}}
}
\date{1 March 2024}
\maketitle      
\thispagestyle{empty}       
\begin{abstract}
\noindent 
We present some basic elements of the theory of generalised Br\`{e}gman relative entropies over nonreflexive Banach spaces. Using nonlinear embeddings of Banach spaces together with the Euler--Legendre functions, this approach unifies two former approaches to Br\`{e}gman relative entropy: one based on reflexive Banach spaces, another based on differential geometry. This construction allows to extend Br\`{e}gman relative entropies, and related geometric and operator structures, to arbitrary-dimensional state spaces of probability, quantum, and postquantum theory. We give several examples, not considered previously in the literature.

\end{abstract}

\section{Introduction}

For any set $Z$, $D:Z\times Z\ra[0,\infty]$ will be called an \df{information} on $Z$ (and $-D$ will be called a \df{relative entropy} on $Z$)\footnote{Cf. \cytat{information is the negative of the quantity (...) defined as entropy} \cite[p. 76]{Wiener:1948}.} if{}f (cf. \cite[p. 1019]{Bregman:1966} \cite[p. 794]{Eguchi:1983} \cite[p. 161]{Csiszar:1995})  $
D(x,y)=0$ $\iff$ $x=y$ $\forall x,y\in Z$. If $\varnothing\neq K\subseteq Z$, $x\in Z$, and $\arginff{y\in K}{D(y,x)}$ (resp., $\arginff{y\in K}{D(x,y)}$) is a singleton set, then we will denote the element of this set by $\LPPP^D_K(x)$ (resp., $\RPPP^D_K(x)$), while the map $x\mapsto\LPPP^D_K(x)$ \cite[p. 32]{Sanov:1957} \cite[Ch. 3.2]{Kullback:1959} (resp., $x\mapsto\RPPP^D_K(x)$ \cite[Eqn. (16)]{Chencov:1968}) will be called a \df{left} (resp., \df{right}) \df{$D$-projection} of $x$ onto $K$.

Let $M$ be a $\mathrm{C}^3$-manifold with a tangent bundle $\T M$, a $\mathrm{C}^3$ riemannian metric tensor $\gbold$ on $\T M$, and a pair $(\nabla,\widetilde{\nabla})$ of $\mathrm{C}^3$ affine connections on $\T M$ (with an arbitrary torsion). Let $\transport^\nabla_c$ denote a $\nabla$-parallel transport in $\T M$ along a curve $c$ in $M$. Then the \df{Norden--Sen geometry} is defined as a quadruple $(M,\gbold,\nabla,\widetilde{\nabla})$ satisfying any of the equivalent conditions \cite[pp. 205--206, \S2, \S4]{Norden:1937} \cite[p. 46]{Sen:1948}:\footnote{In comparison, given $(M,\gbold)$, the Levi-Civita affine connection $\nabla^\gbold$  is characterised among all torsion-free affine connections on $\T M$ by $\gbold(\transport^{\nabla^\gbold}_c(\cdot),\transport^{\nabla^\gbold}_c(\cdot))=\gbold$. Each torsion-free Norden--Sen geometry determines $\nabla^{\gbold}$ by $\nabla^{\gbold}=\frac{1}{2}(\nabla+\widetilde{\nabla})$ \cite[p. 211]{Norden:1937}.}
\begin{align}
\gbold(\transport^{\nabla}_c(\cdot),\transport^{\widetilde{\nabla}}_c(\cdot))&=\gbold,\\
\gbold(\nabla_uv,w)+\gbold(v,\widetilde{\nabla}_uw)&=u(\gbold(v,w))\;\forall u,v,w\in\T M.
\label{eqn.norden.sen}
\end{align}
If $Z$ is a finite dimensional $\mathrm{C}^3$-manifold and $D\in\mathrm{C}^3(Z\times Z;\RR^+)$ has a positive definite hessian matrix, then a third order Taylor expansion of $D$ on $Z$ induces \cite[pp. 795--796]{Eguchi:1983} \cite[p. 357]{Eguchi:1985} a riemannian metric $\gbold^D$ on $\T Z$ and a pair $(\nabla^D,\widetilde{\nabla}^D)$ of torsion-free affine connections on $\T Z$, satisfying the characteristic property \eqref{eqn.norden.sen} of the Norden--Sen geometry. This way the global geometric properties of $D$ can be analysed in local terms of its torsion-free Norden--Sen differential geometry.\footnote{Following \cite[\S4]{Lauritzen:1987:statistical:manifolds}, the torsion-free Norden--Sen geometries are sometimes called ``statistical manifolds''. Apart from not crediting the original authors, this terminology is misleading, since these geometries are independent of any notion of statistics.} 

\section{$D_\Psi$: Br\`{e}gman vs Brunk--Ewing--Utz}

Given a strictly convex, differentiable function $\Psi:\RR^n\ra\RR$ (or $\Psi:M\ra\RR$ with convex $M\subseteq\RR^n$), there are two approaches to construction of a functional encoding the first order Taylor expansion of $\Psi$ (together with its further use in optimisation problems): one going back to Br\`{e}gman's \cite[p. 1021]{Bregman:1966}
\begin{equation}
D_\Psi(x,y):=\Psi(x)-\Psi(y)-\sum\nolimits_{i=1}^n(x_i-y_i)(\grad\,\Psi(y_i))\;\forall x,y\in\RR^n
\label{eqn.Bregman.D.Psi}
\end{equation}
(or $\forall x,y\in M$), another going back to the Brunk--Ewing--Utz \cite[Eqn. (4.4)]{Brunk:Ewing:Utz:1957}
\begin{equation}
D_\Psi^\mu(x,y):=\textstyle\int_{\X\subseteq\RR^m}\mu(\xx)D_\Psi(x(\xx),y(\xx)),
\end{equation}
for $x,y:\X\ra\RR$, $n=1$, and a measure $\mu$ on the Borel subsets of $\RR^m$. 

The former approach has been generalised and widely developed for $\RR^n$ replaced by a reflexive Banach space $(X,\n{\cdot}_X)$ (see Section \ref{section.reflexive}). On the other hand, the latter approach was generalised and further developed for $(\X,\mu)$ given by any countably finite nonzero measure space (see \cite{Csiszar:Matus:2012:kybernetika} and references therein).

The passage from probabilistic to quantum theoretic setting corresponds to replacing $(L_1(\X,\mu),\n{\cdot}_1)$ by the Banach predual $\N_\star$ of a W$^*$-algebra $\N$ (all of these spaces are nonreflexive). The noncommutative analogue $D^{\tr_\H}_\Psi$ of $D_\Psi^\mu$ was  introduced in \cite[\S2.2]{Tsuda:Raetsch:Warmuth:2005} for finite dimensional real Hilbert spaces, and in \cite[pp. 127--129]{Petz:2007}\footnote{More precisely, $D_\Psi^{\tr_\H}(x,y):=\tr_\H(D_\Psi(x,y))$ for a convex and Gateaux differentiable $\Psi:W\ra\BH$, where $W$ is a convex subset of a Banach space, e.g. $W=(\BH)_\star^+$. The evaluation of $D_\Psi^{\tr_\H}(x,y)$ is thus defined by spectral calculus applied to $\Psi$.} for type I W$^*$-algebras (see also \cite[\S{}V]{Harremoes:2017:IEEE} for type I$_n$ JBW-algebras). However, due to nonreflexivity of $\N_\star$, this definition is incapable of utilising the vast body of reflexive Banach space theoretic results obtained for $D_\Psi$, and it is also unclear how to extend the definition of $D_\Psi^{\tr_\H}$ to arbitrary W$^*$-algebras.

For a convex closed $C\subseteq M\subseteq\RR^n$, $D_\Psi$ given by \eqref{eqn.Bregman.D.Psi} exhibits \cite[Lemm. 1]{Bregman:1966}, 
\begin{equation}
D_\Psi(x,\LPPP^{D_\Psi}_C(y))+D_\Psi(\LPPP^{D_\Psi}_C(y),y)\geq D_\Psi(x,y)\;\forall(x,y)\in C\times M
\label{eqn.gen.pyth.left}
\end{equation}
(and analogously for $\RPPP^{D_\Psi}_C$ \cite[Prop. 4.11]{MartinMarquez:Reich:Sabach:2012}; cf. also \cite[Thm. 1]{Chencov:1968}), with $\geq$ replaced by $=$ for affine closed $C$. This property is a nonlinear generalisation of a pythagorean theorem, and is interpreted as an additive decomposition of an (information about) ``data'' into ``signal'' and ``noise''. It is a fundamental feature of $D_\Psi$, characterising $\LPPP^{D_\Psi}_C$ \cite[Cor. 3.35]{Bauschke:Borwein:Combettes:2003} and $\RPPP^{D_\Psi}_C$ \cite[Prop. 4.11]{MartinMarquez:Reich:Sabach:2012}.


\section{$D_\Psi$:  reflexive Banach space setting}\label{section.reflexive}

$(X,\n{\cdot}_X)$ will denote a Banach space over $\RR$. A Banach space $(X^\star,\n{\cdot}_{X^\star})$, consisting of elements given by continuous linear maps $X\ra\RR$, with a norm 
\begin{equation}
\n{y}_{X^\star}:=\sup\{\ab{y(x)}\mid x\in B(X,\n{\cdot}_X):=\{x\in X\mid\n{x}_X\leq1\}\}\;\forall y\in X^\star,
\end{equation}
is called a \df{Banach dual} of $(X,\n{\cdot}_X)$, with respect to a bilinear duality
\begin{equation}
\duality{x,y}_{X\times X^\star}:=y(x)\in\RR\;\forall (x,y)\in X\times X^\star.
\end{equation}
If there exists $(Y,\n{\cdot}_Y)$ with $(Y^\star,\n{\cdot}_{Y^\star})=(X,\n{\cdot}_X)$, then $Y=:X_\star$ is called a \df{predual} of $X$. Symbol $\INT(W)$ (resp., $\cl(W)$) will denote an interior (resp., closure) of $W\subseteq X$ with respect to a topology of $\n{\cdot}_X$.

Given a Banach space $(X,\n{\cdot}_X)$, $\Psi:X\ra\,]-\infty,\infty]$ is called: \df{proper} if{}f 
\begin{equation}
\efd(\Psi):=\{x\in X\mid\Psi(x)\neq\infty\}\neq\varnothing;
\end{equation}
\df{convex} (resp., \df{strictly convex}) if{}f $\forall x,y\in\efd(\Psi)\;\forall\lambda\in\,]0,1[$
\begin{equation}
x\neq y\;\limp\;\Psi(\lambda x+(1-\lambda)y)\leq\mbox{ (resp., }<\mbox{) }\lambda\Psi(x)+(1-\lambda)\Psi(y).
\end{equation}
Let $\pcl(X,\n{\cdot}_X)$ (resp., $\pclg(X,\n{\cdot}_X)$) be the set of all proper, convex, lower semicontinuous functions $\Psi:X\ra\,]-\infty,\infty]$ (resp., that are also Gateaux differentiable on $\intefd{\Psi}\neq\varnothing$, with $\DG\Psi$ denoting a Gateaux derivative of $\Psi$). 

For $\Psi\in\pclg(X,\n{\cdot}_X)$  the \df{Br\`{e}gman function} reads \cite[Eqn. (1)]{Alber:Butnariu:1997} $\forall x\in X$
\begin{equation}
D_\Psi(x,y):=
\Psi(x)-\Psi(y)-\duality{x-y,\DG\Psi(y)}_{X\times X^\star}\;\forall y\in\intefd{\Psi},
\end{equation}
and $D_\Psi(x,y):=\infty$ $\forall y\in X\setminus\intefd{\Psi}$. $D_\Psi$ is an information on $X$ if{}f $\Psi$ is strictly convex on $\intefd{\Psi}$ \cite[Prop. 1.1.9]{Butnariu:Iusem:2000}. 

For a proper $\Psi:X\ra\,]-\infty,\infty]$, a \df{Fenchel dual} map \cite[p. 75]{Fenchel:1949} \cite[p. 8]{Moreau:1962}
\begin{equation}
X^\star\ni y\mapsto\Psi^\lfdual(y):=\sup_{x\in X}\{\duality{x,y}_{X\times X^\star}-\Psi(x)\}\in\,]-\infty,\infty],\label{eqn.Fenchel.duality}
\end{equation}
satisfies $\Psi^\lfdual\in\pcl(X^\star,\n{\cdot}_{X^\star})$ \cite[Thm. 3.6]{Broendsted:1964}. If $(X,\n{\cdot}_X)$ is reflexive and $\Psi\in\pclg(X,\n{\cdot}_X)$, then $\Psi$ will be called \df{Euler--Legendre}\footnote{These functions are usually called ``Legendre'' (for $X=\RR^n$ they were introduced namelessly in \cite[Thm. C-K]{Rockafellar:1963}). Yet, the transformation $\dd(z(x,y)-px-qy)=-x\dd p-y\dd q$, with $p=\frac{\partial z(x,y)}{\partial x}$ and $q=\frac{\partial z(x,y)}{\partial y}$, was introduced first by Euler \cite[Part I, Probl. 11]{Euler:1770}, and only 17 years later by Legendre \cite[p. 347]{Legendre:1787}.} if{}f \cite[Def. 5.2.(iii), Thm. 5.4, Thm. 5.6]{Bauschke:Borwein:Combettes:2001} \cite[\S2.1]{Reich:Sabach:2009} $\Psi^\lfdual\in\pclg(X^\star,\n{\cdot}_{X^\star})$ and
\begin{equation}
\left\{\begin{array}{l}
\efd(\DG\Psi):=\{x\in\efd(\Psi)\mid\exists\;\DG\Psi(x)\}=\intefd{\Psi},\\
\efd(\DG\Psi^\lfdual)=\intefd{\Psi^\lfdual}.
\end{array}\right.
\end{equation}

For $X=\RR^n$, the definition of Euler--Legendre functions goes back to Rockafellar, who showed \cite[Thm. C-K]{Rockafellar:1963} \cite[Thm. 1]{Rockafellar:1967} that if $\varnothing\neq U\subseteq\RR^n$ is open and convex, while $\Psi:U\ra\,]-\infty,\infty]$ is strictly convex, differentiable on $U$, and
\begin{equation}
\lim_{t\ra^+0}\textstyle\frac{\dd}{\dd t}\Psi(tx+(1-t)y)=-\infty\;\;\forall(x,y)\in U\times(\cl(U)\setminus U),
\end{equation}
then $\grad\,\Psi$ is a bijection on $U$, $\grad(\Psi^\lfdual)=(\grad\,\Psi)^{-1}$ on $(\grad\,\Psi)(U)$, and $\Psi^\lfdual$ on $(\grad\,\Psi)(U)$ satisfies the same conditions as $\Psi$ on $U$.

\section{$D_\Psi$: dually flat setting}

The \df{dually flat} (a.k.a. \df{hessian}) geometry \cite[Prop. (p. 213)]{Shima:1976} is characterised among all torsion-free Norden--Sen geometries by the flatness of $\nabla$ and $\widetilde{\nabla}$. This is equivalent with existence of two coordinate systems, $\{\theta_i\mid i\in\{1,\ldots,n\}\}:M\ra\RR^n$ and $\{\eta_i\mid i\in\{1,\ldots,n\}\}:M\ra\RR^n$, such that, $\forall\rho\in M$, 
\begin{empheq}[left={ \empheqlbrace}]{align}
\eta_i(\rho)&=\frac{\partial\Psi(\theta(\rho))}{\partial\theta^i},\;\;\theta_i(\rho)=\frac{\partial\Psi^\lfdual(\eta(\rho))}{\partial\eta^i}\label{eqn.eta.theta.Phi}\\
\Psi^\lfdual(y)&=\sup_{x\in\RR^n}\left\{\sum_{i=1}^nx_iy_i-\Psi(x)\right\}\; \forall x\in\RR^n,\label{eqn.finite.Fenchel.duality}
\end{empheq}
and, for $D_{\theta,\Psi}(\rho,\sigma):=D_\Psi(\theta(\rho),\theta(\sigma))$ with $D_\Psi$ defined by \eqref{eqn.Bregman.D.Psi},  
\begin{empheq}[left={ \empheqlbrace}]{align}
\Upgamma^{\nabla^{D_{\theta,\Psi}}}_{ijk}(\theta(\rho))&=0,\;\;\Upgamma^{\widetilde{\nabla}^{D_{\eta,\Psi}}}_{ijk}(\eta(\rho))=0\label{eqn.upgamma.nabla}\\
\gbold^{D_{\theta,\Psi}}_{ij}(\theta(\rho))&=\frac{\partial^2\Psi(\theta(\rho))}{\partial\theta^i\partial\theta^j},
\end{empheq}
where $\Upgamma^\nabla(u,v,w):=\gbold(\nabla_uv,w)$ $\forall u,v,w\in\T M$, while the subscript ${}_i$ denotes evaluation at the $i$-th component of a basis in $\T M$ given by coordinate system differentials (i.e., setting $u=\frac{\partial}{\partial\theta^i}$, etc., in \eqref{eqn.upgamma.nabla}). (Also, $\gbold^{D_{\eta,\Psi}}_{ij}(\eta(\rho))=\frac{\partial^2\Psi^\lfdual(\eta(\rho))}{\partial\eta^i\partial\eta^j}$.) When reconsidered in this setting, the left (resp., right) generalised py\-tha\-go\-re\-an theorem is equivalent with: a projection of $y\in M$ onto $C$ along $\widetilde{\nabla}^{D_{\eta,\Psi}}$-(resp., $\nabla^{D_{\theta,\Psi}}$-)geodesics is $\gbold^{D_{\theta,\Psi}}$-orthogonal (= $\gbold^{D_{\eta,\Psi}}$-orthogonal) to $C$ \cite[Thm. 3.4]{Amari:Nagaoka:1993}.

Equation \eqref{eqn.finite.Fenchel.duality} is a special case of \eqref{eqn.Fenchel.duality}. Furthermore, \eqref{eqn.eta.theta.Phi} require only $\mathrm{C}^1$-differentiability. The approach presented in Section \ref{section.generalised.Bregman} is rooted in an observation that the correct generalisation of \eqref{eqn.eta.theta.Phi} requires two components: Euler--Legendre $\Psi$ on a reflexive Banach space $(X,\n{\cdot}_X)$, and nonlinear embeddings into $(X,\n{\cdot}_X)$ and $(X^\star,\n{\cdot}_{X^\star})$, replacing, respectively, $\theta$ and $\eta$.

\section{$D_{\ell,\Psi}$}\label{section.generalised.Bregman}

In \cite[\S3]{Kostecki:2017} we introduced a generalisation, $D_{\ell,\Psi}$, of a family of Br\`{e}gman informations $D_\Psi$ on reflexive Banach spaces $(X,\n{\cdot}_X)$,  applicable to a wide range of nonreflexive Banach spaces $(Y,\n{\cdot}_Y)$. (E.g., to postquantum state spaces, given by bases $Z\subseteq V^+$ of positive cones $V^+$ of radially compact base normed spaces in spectral duality, $(V,\n{\cdot}_V)=(Y,\n{\cdot}_Y)$.) The main idea is to pull back the properties exhibited by $D_\Psi$ with Euler--Legendre $\Psi$ acting on $(X,\n{\cdot}_X)$ into the properties exhibited by $D_{\ell,\Psi}(\cdot,\cdot):=D_\Psi(\ell(\cdot),\ell(\cdot))$, where $\ell:Z\ra X$ and $Z\subseteq Y$.

\begin{definition}\label{def.ell.Psi.information}
\textup{\cite[Def. 3.1]{Kostecki:2017}} Let $(Y,\n{\cdot}_Y)$ be a Banach space, let $(X,\n{\cdot}_X)$ be a reflexive Banach space, let $\Psi\in\pclg(X,\n{\cdot}_X)$ be strictly convex on $\intefd{\Psi}$ and Euler--Legendre, let $\varnothing\neq Z\subseteq Y$, and let $\ell:Z\ra\ell(Z)\subseteq X$ be a bijection such that $\ell(Z)\cap\intefd{\Psi}\neq\varnothing$. Then:
\begin{enumerate}
\item[(i)] if $\varnothing\neq C\subseteq Y$, and $\ell(C)$ is convex (resp., closed; affine), then $C$ will be called \df{$\ell$-convex} (resp., \df{$\ell$-closed}; \df{$\ell$-affine}); 
\item[(ii)] a triple $(Z,\ell,\Psi)$ will be called a \df{generalised pythagorean geometry};
\item[(iii)] an \df{$(\ell,\Psi)$-information} (a \df{generalised Br\`{e}gman information}) on Z is
\begin{equation}
D_{\ell,\Psi}(\phi,\psi):=D_\Psi(\ell(\phi),\ell(\psi))
\;\;
\forall(\phi,\psi)\in Z\times\ell^{-1}(\ell(Z)\cap\intefd{\Psi}).
\label{eqn.generalised.Bregman}
\end{equation}
\end{enumerate}
\end{definition}
\begin{proposition}\label{prop.D.ell.psi.properties}
\textup{\cite[Prop. 3.2]{Kostecki:2017}} Under assumptions of Definition \ref{def.ell.Psi.information}, let $\varnothing\neq C\subseteq Z$ be $\ell$-convex and $\ell$-closed, and let $\psi\in\ell^{-1}(\ell(Z)\cap\intefd{\Psi})$. Then: 
\begin{enumerate}
\item[(i)] $D_{\ell,\Psi}$ is an information on $Z$;
\item[(ii)] $\arginff{\phi\in C}{D_{\ell,\Psi}(\phi,\psi)}$ is a singleton set, denoted $\{\LPPP^{D_{\ell,\Psi}}_C(\psi)\}$;
\item[(iii)] $\omega\in C$ is the unique solution of $D_{\ell,\Psi}(\phi,\omega)+D_{\ell,\Psi}(\omega,\psi)\leq D_{\ell,\Psi}(\phi,\psi)$ $\forall\phi\in C$ if{}f $\omega=\LPPP^{D_{\ell,\Psi}}_C(\psi)$ (in `then' case, if $C$ is $\ell$-affine, then $=$ replaces $\leq$);
\item[(v)] if $\ell$ is norm-to-norm continuous and $\LPPP^{D_{\Psi}}_K$ is norm-to-norm continuous for any convex closed $\varnothing\neq K\subseteq\ell(Z)\cap\intefd{\Psi}$, then $\LPPP^{D_{\ell,\Psi}}_C$ is norm-to-norm continuous for any $\ell$-convex and closed $\varnothing\neq C\subseteq\ell^{-1}(\ell(Z)\cap\intefd{\Psi})$.
\end{enumerate}
\end{proposition}
An analogous result for $\RPPP^{D_{\ell,\Psi}}$ also holds \cite[Part I]{Kostecki:2023:I:II:III:IV} (cf. also \cite[Thm. 1]{Chencov:1968}). 

For $X=\RR^n$, $D_{\ell,\Psi}$ recovers the setting of Br\`{e}gman information $D_{\theta,\Psi}$ on an $n$-dimensional $\mathrm{C}^1$-manifold (hence, in particular, $\mathrm{C}^\infty$-manifold) $M$, with the map $\ell:M\ra\RR^n$ (resp., $\DG\Psi\circ\ell:M\ra\RR^n$) given by the coordinate system $\{\theta_i\}$ (resp., $\{\eta_i\}$). More specifically, a domain $M$ of a dually flat geometry is assumed to be a (suitably differentiable) manifold, covered by two global maps $\{\theta_i\}$ and $\{\eta_i\}$, \textit{without} assuming $M\subseteq\RR^n$, cf. \cite{Amari:Nagaoka:1993,Shima:2007}. This is not addressed by \eqref{eqn.Bregman.D.Psi}, and is addressed (up to a weaker assumption on the order of differentiability) by \eqref{eqn.generalised.Bregman}. 

This way the framework of generalised Br\`{e}gman information $D_{\ell,\Psi}$ unifies reflexive Banach space theoretic and finite dimensional smooth information geometric approaches to Br\`{e}gman information. If $\ell$ is a norm-to-norm continuous homeomorphism, then the $\ell$-closed sets in $Z$ are closed in terms of topology of $\n{\cdot}_Y$. This fragment of a theory provides a fusion of nonlinear convex analysis with nonlinear homeomorphic theory of Banach spaces. In particular, if $\ell$ is H\"{o}lder continuous, then it allows to pull back the conditions on H\"{o}lder continuity of $\LPPP^{D_{\Psi}}_K$ and $\RPPP^{D_{\Psi}}_K$ into results on H\"{o}lder continuity of $\LPPP^{D_{\ell,\Psi}}_C$ and $\RPPP^{D_{\ell,\Psi}}_C$. Generalised pythagorean geometry $(Z,\ell,\Psi)$ is a more general object than $D_{\ell,\Psi}$, and allows to suitably generalise also the affine connections \eqref{eqn.upgamma.nabla} \cite[Part IV]{Kostecki:2023:I:II:III:IV}.
 
In this context, our approach arises partially from an observation that the $\ell_\gamma$ (resp., $\ell_\orlicz$) embeddings, cf. Example \hyperlink{target.Breg.example.61a}{6.1.(a)} (resp., \hyperlink{target.Breg.example.61c}{6.1.(c)}) below, used in \cite[Eqn. (2.7)]{Nagaoka:Amari:1982} (resp., \cite[\S7.2]{Gibilisco:Pistone:1998}), are finite dimensional Mazur (resp., Kaczmarz) maps \cite[p. 83]{Mazur:1929} (resp., \cite[p.148]{Kaczmarz:1933}) on $(L_1(\X,\mu))^+$. Drawing from an important example in \cite[\S6--\S8]{Jencova:2005} (equal to Example \hyperlink{target.Breg.example.61a}{6.1.(a)} with $\alpha=\gamma(1-\gamma)$ and $\beta=\gamma$), an abstract framework aiming at this unification was proposed in \cite[Eqns. (24), (31)]{Kostecki:2011:OSID}, while its implementation, based on the use of Euler--Legendre $\Psi$, was given in \cite[\S3--\S4]{Kostecki:2017}. The resulting theory is developed in details in \cite{Kostecki:2023:I:II:III:IV}.

\section{Examples of $(\ell,\Psi)$ with $Z\subseteq V^+$ (for Proposition \ref{prop.D.ell.psi.properties})} 

If $(Y,\n{\cdot}_Y)$ is partially ordered by $\geq$, then $Y^+:=\{x\in Y\mid x\geq0\}$. All examples below feature $(Y,\n{\cdot}_Y)$ given by some kind of a radially compact base normed space $(V,\n{\cdot}_V)$. Such spaces provide the setting for the (linear) convex operational generalisation of quantum theory (a.k.a. ``generalised probability theory'' or ``postquantum theory''), with state space given by $V_1^+:=\{\phi\in V^+\mid\n{x}_V=1\}$.

\hypertarget{target.Breg.example.61}{}\subsection*{Example 6.1.}\label{example.six.one}
\begin{enumerate}
\item[\hypertarget{target.Breg.example.61a}{\textit{(a)}}.] (=\cite[Prop. 4.2]{Kostecki:2017}.) If $\N$ is a W$^*$-algebra, $\alpha\in\,]0,\infty[$, $\beta,\gamma\in\,]0,1[$, $(X,\n{\cdot}_X)=(L_{1/\gamma}(\N),\n{\cdot}_{1/\gamma})$, then the Mazur map
\begin{equation}
\ell=\ell_\gamma:Z=\N_\star^+\ni\phi\mapsto\phi^\gamma\in (L_{1/\gamma}(\N))^+
\end{equation}
is H\"{o}lder continuous \cite[Thm. (p. 37)]{Ricard:2015}. If $\Psi=\Psi_{\alpha,\beta}:=\frac{\beta}{\alpha}\n{\cdot}_{X}^{1/\beta}$, then
\begin{equation}
D_{\ell_\gamma,\Psi_{\alpha,\beta}}(\phi,\psi)=\alpha^{-1}(\beta\n{\phi}^{\gamma/\beta}_1+(1-\beta)\n{\psi}_1^{\gamma/\beta}-\n{\psi}_1^{\gamma/\beta-1}\textstyle\int(\phi^\gamma\psi^{1-\gamma}))
\end{equation}
$\forall\phi,\psi\in\N_\star^+$, where $\int$ is understood as in \textup{\cite[Eqn. (3.12')]{Falcone:Takesaki:2001}}; if $\N=\BH:=\{$bounded operators on a Hilbert space $\H\}$, then $\N_\star=\schatten_1(\H)\equiv\{$trace class operators on $\H\}$, $L_{1/\gamma}(\N)=:\schatten_{1/\gamma}(\H)$, and $\int\cdot=\tr_\H(\cdot)=\n{\cdot}_1$. 
\item[\textit{(b)}.] (=\cite[Prop. 4.7]{Kostecki:2017}.) Let $A$ be a semifinite JBW-algebra with a Jordan product $\bullet$, a faithful normal semifinite trace $\tau$, $\alpha\in\,]0,\infty[$, $\beta,\gamma\in\,]0,1[$, $(X,\n{\cdot}_X)=(L_{1/\gamma}(A,\tau),\n{\cdot}_{1/\gamma})$, $\Psi=\Psi_{\alpha,\beta}$. Then $\ell=\ell_\gamma:A_\star^+\ni\phi\mapsto\phi^\gamma\in(L_{1/\gamma}(A,\tau))^+$ is H\"{o}lder continuous \cite[Prop. 4.6]{Kostecki:2017}, and $\forall\omega,\phi\in Z=A_\star^+$ $D_{\ell_\gamma,\Psi_{\alpha,\beta}}(\omega,\phi)=$
\begin{equation}
\alpha^{-1}(\beta(\tau(\omega))^{\gamma/\beta}+(1-\beta)(\tau(\phi))^{\gamma/\beta}-(\tau(\phi))^{\gamma/\beta-1}\tau(\omega^\gamma\bullet\phi^{1-\gamma})).
\end{equation}
\item[\hypertarget{target.Breg.example.61c}{\textit{(c)}}.] (=\cite[Cor. 4.12]{Kostecki:2017}.) If $(\X,\mu)$ is a nonatomic measure space, $\mu(\X)<\infty$, $\orlicz:\RR\ra\RR^+$ is even, strictly convex, continuously differentiable, with $\orlicz(1)=1$, $\orlicz(u)=0$ if{}f $u=0$, $\limsup_{u\ra\infty}\frac{\orlicz(2u)}{\orlicz(u)}<\infty$, $\liminf_{u\ra\infty}\frac{\orlicz(2u)}{\orlicz(u)}>2$, $\lim_{u\ra^+0}\frac{\orlicz(u)}{u}=0$, $\lim_{u\ra\infty}\frac{\orlicz(u)}{u}=\infty$, $t,s\in\RR^+$, $t<s$, $u\mapsto\frac{\orlicz^{-1}(u)}{u^t}$ is nondecreasing, and $u\mapsto\frac{\orlicz^{-1}(u)}{u^s}$ is nonincreasing, then the Kaczmarz map
\begin{equation}
\ell=\ell_\orlicz:Z=(L_1(\X,\mu))^+_1\ni\phi\mapsto\orlicz^{-1}(\phi)\in(L_\orlicz(\X,\mu))^+_1
\end{equation}
is H\"{o}lder continuous for the Morse--Transue--Nakano--Luxemburg norm $\n{\cdot}_{\orlicz}$ on Orlicz space $L_\orlicz(\X,\mu)$ \cite[Cor. 2.5]{Delpech:2005}. For $\Psi=\Psi_{\beta,\beta}$, $\beta\in\,]0,1[$, this gives
\begin{equation}
D_{\ell_\orlicz,\Psi_{\beta,\beta}}(\omega,\phi)=\beta^{-1}(1-\bar{\orlicz}(\omega,\phi)/\bar{\orlicz}(\phi,\phi)),
\end{equation}
where $\bar{\orlicz}(\omega,\phi):=\int\mu\orlicz^{-1}(\omega)\orlicz'(\orlicz^{-1}(\phi))$, and $(\cdot)'$ denotes a derivative.
\end{enumerate}
All these cases have norm-to-norm continuous $\LPPP^{D_{\ell,\Psi}}_C$. In \cite{Kostecki:2023:I:II:III:IV} we prove this also for $\RPPP^{D_{\ell,\Psi}}_C$, and establish conditions for H\"{o}lder continuity of $\LPPP^{D_{\ell,\Psi}}_C$ and $\RPPP^{D_{\ell,\Psi}}_C$.
\subsection*{Example 6.2.}\label{example.six.two}

(= \cite[Prop. 4.14]{Kostecki:2017} for $\varphi(t)=\varphi_{\alpha,\beta}(t)=\frac{1}{\alpha}t^{1/\beta-1}$, i.e. $\Psi=\Psi_{\alpha,\beta}=\Psi_{\varphi_{\alpha,\beta}}$; \cite[Part I]{Kostecki:2023:I:II:III:IV} for $\Psi=\Psi_\varphi$). Let $(V,\n{\cdot}_V)$ be a generalised spin factor \cite[Def. 4]{Berdikulov:Odilov:1995}, i.e. $V=\RR\oplus X$, where $(X,\n{\cdot}_X)$ is a reflexive Banach space, and
\begin{equation}
\forall v=(\lambda,x)\in V\;\;\left\{\begin{array}{l}
v\geq0\;:\iff\;\lambda\geq\n{x}_X\\
\n{v}_V:=\max\{\ab{\lambda},\n{x}_X\}.
\end{array}\right.
\end{equation}
Let $\Psi(x)=\Psi_\varphi(x):=\int_0^{\n{x}_X}\dd t\,\varphi(t)$, where $\varphi:\RR^+\ra\RR^+$ is positive, strictly increasing, continuous, $\varphi(0)=0$, and $\lim_{t\ra\infty}\varphi(t)=\infty$.\footnote{Cf. \cite[Rem. 4.15]{Kostecki:2017}. In \cite{Kostecki:2023:I:II:III:IV} we also extend Example \hyperlink{target.Breg.example.61}{6.1} to $\Psi=\Psi_\varphi$.}
 Then $\Psi_\varphi$ (and, in particular, $\Psi_{\alpha,\beta}$) is Euler--Legendre if{}f $(V,\n{\cdot}_V)$ satisfies spectral duality condition \cite[Def. (p. 55)]{Alfsen:Shultz:1976}. This gives a family $D_{\ell_X,\Psi_\varphi}$ on $Z=\{w\in V^+\mid \n{w}_V=1\}$, where
\vspace{-0.4cm}
\begin{equation}
\ell=\ell_X:Z\ni v=:(1,x)\mapsto x\in B(X,\n{\cdot}_X).
\end{equation}
\vspace{-0.9cm}
\subsection*{Example 6.3.}\label{example.six.three}

Let $\H$ be a Hilbert space over $\CC$ with $n:=(\dim\H)^2\in\NN$ (hence, $\schatten_{1/\widetilde{\gamma}}(\H)=\schatten_{1/\gamma}(\H)$ $\forall\gamma,\widetilde{\gamma}\in\,]0,1[$). Let $(\,\cdot\,)^\sa$ := self-adjoint part of $(\,\cdot\,)$. Let $\lewis(x)$, with 
\begin{equation}
\K:=(\schatten_2(\H))^\sa=\{\mbox{hermitean }n\times n\mbox{ matrices}\}\ni x\mapsto\lewis(x)\in\RR^n,
\end{equation}
be a vector of eigenvalues of $x$ ordered nonincreasingly. For $\Phi:\RR^n\ra\,]-\infty,\infty]$, let $\Phi(s(x))=\Phi(x)$ $\forall$ permutation matrices $s:\RR^n\ra\RR^n$. Then $\Psi=\Phi\circ\lewis$ is Euler--Legendre if{}f $\Phi$ is Euler--Legendre \cite[Cor. 3.2, Cor. 3.3]{Lewis:1996:hermitian}. E.g., if: $\Phi(x)=$ 
\begin{enumerate}
\item[\textit{(a)}.] \cite[Ex. 6.5, Cor. 5.13]{Bauschke:Borwein:1997} $\sum_{i=1}^n(x_i\log(x_i)-x_i)$ if $x\geq0$, and $\infty$ otherwise; 
\item[\textit{(b)}.] \cite{Burg:1967} \cite[Ex. 6.7, Cor. 5.13]{Bauschke:Borwein:1997} $-\sum_{i=1}^n\log(x_i)$ on $\,]0,\infty[^n$, and $\infty$ otherwise;\footnote{$D_\Phi(x,y)=\sum\nolimits_{i=1}^n(-\log\frac{x_i}{y_i}+\frac{x_i}{y_i}-1)$ $\forall(x,y)\in(\RR^n)^+_0\times(\RR^n)^+_0$, corresponding to $\Phi$ in (b), was introduced by Pinsker in \cite[Eqn. (4)]{Pinsker:1960:DAN} \cite[Eqn. (10.5.4)]{Pinsker:1960}. The result by Itakura--Saito \cite[Eqn. (7)]{Itakura:Saito:1968}, usually cited as a reference for this $D_\Phi$, has appeared 8 years later, and contains only a formula $2\log(2\pi)+\frac{1}{2\pi}\int_{-\pi}^{\pi}\dd t(\log(y(t))+\frac{x(t)}{y(t)})$.}
\item[\textit{(c)}.] \cite[Eqn. (60)]{Kapur:1972} \cite[Ex. 6.6, Cor. 5.13]{Bauschke:Borwein:1997} $\sum_{i=1}^n(x_i\log(x_i)+(1-x_i)\log(1-x_i))$ on $[0,1]^n$, and $\infty$ otherwise; 
\item[\textit{(d)}.] \cite[Ex. 6.1, Cor. 5.13]{Bauschke:Borwein:1997} $\sum_{i=1}^n\gamma\ab{x_i}^{1/\gamma}$ on $\RR^n$ with $\gamma\in\,]0,1[$; 
\item[\textit{(e)}.] \cite[Eqn. (37)]{Reem:Reich:DePierro:2019} \cite[\S7.2]{Reem:Reich:DePierro:2019} $\Phi_\alpha(x):=\frac{1}{\alpha-1}\sum_{i=1}^n(x_i^\alpha-1)$ for $(x,\alpha)\in[0,\infty[^n\times\,]0,1[$, $-\Phi_\alpha(x)$ for $(x,\alpha)\in\,]0,\infty[^n\times\,]-\infty,0[$, and $\infty$ otherwise;\footnote{Cf.: $-\frac{2^{\alpha-1}(\alpha-1)}{2^{\alpha-1}-1}(\Phi_\alpha+\frac{n-1}{\alpha-1})$ $\forall\alpha>0$ in \cite[Thm. 1]{Havrda:Charvat:1967}; $-\Phi_\alpha-\frac{n-1}{\alpha-1}$ $\forall\alpha\in\RR$ in \cite[Eqn. (1)]{Tsallis:1988}; a detailed analysis when $\frac{1}{\alpha}(-\Phi_\alpha-\frac{n}{\alpha-1})$ is Euler--Legendre in \cite[Thm. 5]{Woo:2017}.}
\end{enumerate}
and $\K_0^+:=(\schatten_2(\H))_0^+=\{$strictly positive definite $n\times n$ matrices$\}$, then: $D_{\Phi\circ\lewis}(\xi,\zeta)=$ 
\begin{enumerate}
\item[\textit{(a)}.] \cite[Def.1]{Umegaki:1961} $\tr_\H(\xi(\log\xi-\log\zeta)-\xi-\zeta)$ $\forall(\xi,\zeta)\in\K^+\times\K^+_0$; 
\item[\textit{(b)}.] \cite[\S5]{James:Stein:1961} $\s{\xi,\zeta^{-1}}_{\K}-\log\det(\xi\zeta^{-1})-n=h(\zeta^{-1/2}\xi\zeta^{-1/2})-n$ $\forall(\xi,\zeta)\in\K^+_0\times\K^+_0$, for $h(\xi):=\tr_{\K}(\xi)-\log\det(\xi)$; 
\item[\textit{(c)}.] \cite[p. 376]{Nock:Magdalou:Briys:Nielsen:2013} $\tr_\H(\xi(\log\xi-\log\zeta)+(\II-\xi)(\log(\II-\xi)-\log(\II-\zeta)))$ $\forall(\xi,\zeta)\in B^+\times\INT(B^+)$, where $B^+:=\K^+\cap B(\K,\n{\cdot}_2)$; 
\item[\textit{(d)}.] \cite[Cor. 4.18.(ii)]{Kostecki:2017} $\tr_\H(\gamma\ab{\xi}^{1/\gamma}+(1-\gamma)\zeta^{1/\gamma}-\xi\zeta^{1/\gamma-1})$ $\forall(\xi,\zeta)\in\K\times\K_0^+$ (under restriction of a domain of $\zeta$ to $\K_0^+$);
\item[\textit{(e)}.] \cite[Cor. 4.18.(iii)]{Kostecki:2017} $D_{\alpha}(\xi,\zeta):=\tr_\H(\zeta^\alpha-\frac{1}{1-\alpha}\xi^\alpha+\frac{\alpha}{1-\alpha}\zeta^{\alpha-1}\xi)$ $\forall(\xi,\zeta,\alpha)\in\K^+\times\K^+_0$ $\times\,]0,1[$, $-D_\alpha(\xi,\zeta)$ $\forall(\xi,\zeta,\alpha)\in\K^+_0\times\K^+_0\times\,]-\infty,0[$;
\end{enumerate}
with ``$D_{\Phi\circ\lewis}(\xi,\zeta):=\infty$ otherwise'' in all cases, and $\s{\xi,\zeta}_{\K}:=\tr_{(\schatten_2(\H))^\sa}(\xi\zeta)$. 
 All cases (a)--(e) of $D_{\Phi\circ\lewis}$ are also the special cases of $D_\Psi^{\tr_\H}$, with a range of good optimisation theoretic properties implied by the fact that $\Phi\circ\lewis$ is Euler--Legendre. $\ell$ can be set to be any automorphism of $(\schatten_2(\H))^\sa$ preserving $\intefd{\Phi\circ\lewis}$, e.g. a restriction of $\ell_{1/2}$ to a subset of $(\schatten_1(\H))^\sa$, corresponding to $\intefd{\Phi\circ\lewis}$.

\subsubsection*{Acknowledgements} I thank: Lucien Hardy, Ravi Kunjwal, Jerzy Lewandowski, and Marcin Marciniak for hosting me as a visitor;  Fran\-ce\-sco Buscemi, Paolo Gibilisco, and Anna Jen\v{c}ov\'{a} for hospitality and discussions;  Micha{\l} Eckstein, Jan G{\l}owacki, and Karol Horodecki for help; Perimeter Institute for Theoretical Physics and Polish National Science Centre (grants 2015/18/\allowbreak{}E/ST2/00327 and 2021/42/A/ST2/00356) for support. Research at Perimeter Institute is supported by the Government of Canada through Industry Canada and by the Province of Ontario through the Ministry of Research and Innovation.

\vspace{0.6cm}\begin{spacing}{0.9}\noindent{\small\textit{Remark.} Cyrillic names and titles were bijectively transliterated from the original, using the system: {\cyrrm{ts}} = c, {\cyrrm{ch}} = ch, {\cyrrm{kh}} = kh, {\cyrrm{zh}} = zh, {\cyrrm{sh}} = sh, {\cyrrm{shch}} = \v{s}, {\cyrrm{i}} = i, {\cyrrm{\u{i}}} = \u{\i}, i = \={\i}, {\cyrrm{y}} = y, {\cyrrm{yu}} = yu, {\cyrrm{ya}} = ya, {\cyrrm{\"{e}}} = \"{e}, {\cyrrm{\`{e}}} = \`{e}, {\cyrrm{\cdprime}} = `, {\cyrrm{\cprime}} = ', and analogously for capitalised letters. Symbol * in front of a bibliographic item indicates that I have not seen this work.}\end{spacing}

\section*{References}
\addcontentsline{toc}{section}{References}
{\small
\begingroup
\raggedright
\bibliographystyle{rpkbib}
\renewcommand\refname{\vskip -1cm}

\bibliography{rpk}  
\endgroup        
}
\end{document}